\documentclass[a4paper,11pt]{article}
\usepackage{pos}

\usepackage{multirow, lscape}
\usepackage{multicol}
\usepackage{array}
\usepackage{slashed}
\usepackage{graphicx}
\usepackage{amsthm,amsmath}

\usepackage{amssymb}
\usepackage{mathrsfs}
\usepackage{bm}
\usepackage{pifont}
\usepackage{tcolorbox}
\usepackage{colortbl}
\usepackage{chngpage}
\usepackage[export]{adjustbox}

\usepackage{url}
\begin{document}

\title{Gluon Parton Distribution of the Pion and Nucleon from Lattice QCD}

\ShortTitle{Gluon Parton Distribution of the Pion and Nucleon from Lattice QCD}

\author*[a]{Zhouyou Fan}
\author[a,b]{Huey-Wen Lin}
\affiliation[a]{Department of Physics and Astronomy, Michigan State University, MI, 48824, U.S.A}
\affiliation[b]{Department of Computational Mathematics, Science and Engineering, Michigan State University, MI, 48824, U.S.A}
\emailAdd{fanzhouy@msu.edu}

\abstract{
We present the $x$-dependent nucleon and pion gluon distribution from lattice QCD using the pseudo-PDF approach, on lattice ensembles with $2+1+1$ flavors of highly improved staggered quarks (HISQ), generated by MILC Collaboration. We use clover fermions for the valence action and momentum smearing to achieve pion boost momentum up to 2.56~GeV on three lattice spacings $a\approx 0.9, 0.12$ and 0.15~fm and three pion masses $M_{\pi}\approx 220$, 310 and 690~MeV.
We compare our pion and preliminary nucleon gluon results with the determination by global fits.}

\FullConference{The 38th Annual International Symposium on Lattice Field Theory - LATTICE2021\\
		Massachusetts Institute of Technology, Cambridge, Massachusetts, USA.\\
		26-30 July, 2021}

\maketitle

\section{Introduction}
Parton distribution functions (PDFs) are important to characterize the structure of the hadron and nonperturbative QCD. There have been many detailed studies of quark structure of nucleon and pion during the past few decades. Gluon structure is also important.
The gluon PDF dominates at small $x$, and its error at large $x$ is large compared to the valence-quark PDFs.
Current gluon PDFs from different global analyses vary by the input experimental data and fit strategies.
Gluon nucleon and pion PDFs are mostly studied by global analysis of experimental data~\cite{Novikov:2020snp,Barry:2018ort,Cao:2021aci,Hou:2019efy,NNPDF:2017mvq}. Theoretically, lattice QCD is an independent approach to calculate the gluon PDF.

In this talk, we present our calculation of pion and nucleon gluon PDFs on clover valence fermions on four ensembles with $N_f = 2+1+1$ highly improved staggered quarks (HISQ)~\cite{Follana:2006rc} generated by the MILC Collaboration~\cite{Bazavov:2010ru,Bazavov:2012xda,Bazavov:2017lyh,FermilabLattice:2017wgj} with three different lattice spacings ($a\approx 0.9, 0.12$ and 0.15~fm) and three pion masses (220, 310, 690~MeV), as shown in Table.~\ref{table-data}.
Following the study in Ref.~\cite{Fan:2018dxu}, five HYP-smearing~\cite{Hasenfratz:2001hp} steps are used on the gluon loops to reduce the statistical uncertainties.
The measurements vary $10^5$--$10^6$ for different ensembles. More measurements and various lattice spacings are studied comparing to our previous nucleon gluon PDF calculation on a12m310 ensemble~\cite{Fan:2020cpa}.
In Sec.~\ref{sec:cal-details}, we present the pseudo-PDF procedure to obtain the lightcone gluon PDF and how we extracted the reduced pseudo Ioffe-time distribution (pITDs) from lattice calculated correlators.
In Sec.~\ref{sec:results}, the final determination of the gluon pion and nucleon PDFs from our lattice calculations is compared with the phenomenology global fit PDF results.
A discussion of the systematics and the outlook for the gluon pion and nucleon PDFs are included in the last Sec.~\ref{sec:summary}.

\begin{table}[!htbp]
\centering
\begin{tabular}{|c|c|c|c|c|}
\hline
  ensemble & a09m310 & a12m220 & a12m310 & a15m310 \\
\hline
  $a$ (fm) & $0.0888(8)$ & $0.1184(10)$ & $0.1207(11)$  & $0.1510(20)$ \\
\hline
  $L^3\times T$ & $32^3\times 96$ & $32^3\times 64$ & $24^3\times 64$ & $16^3\times 48$ \\
\hline
  $M_{\pi}^\text{val}$ (MeV) & $313.1(13)$ & $226.6(3)$ & $309.0(11)$ & $319.1(31)$\\
\hline
  $M_{\eta_s}^\text{val}$ (MeV) & 698.0(7) & $696.9(2)$ & $684.1(6)$ & $687.3(13)$\\
\hline
  $P_z$ (GeV)  & [0,2.18] & $[0,2.29]$ &  $[0,2.14]$ &  $[0,2.56]$   \\
\hline
  $N_\text{meas}^\text{pion}$ & N/A &  731,200  & 143,680 & 21,600  \\
\hline
  $N_\text{meas}^\text{nucleon}$ & 145,296 &  N/A  & 324,160 & 21,600  \\
\hline
  $t_\text{sep}$ & \{6,7,8,9\}& \{5,6,7,8\}   & \{6,7,8,9\} & \{6,7,8,9\}  \\
\hline
\end{tabular}
\caption{
Lattice spacing $a$, valence pion mass $M_\pi^\text{val}$ and $\eta_s$ mass $M_{\eta_s}^\text{val}$, lattice size $L^3\times T$, number of configurations $N_\text{cfg}$, number of total two-point correlator measurements $N_\text{meas}^\text{2pt}$, and separation times $t_\text{sep}$ used in the three-point correlator fits of $N_f=2+1+1$ clover valence fermions on HISQ ensembles generated by the MILC collaboration and analyzed in this study.
}
\label{table-data}
\end{table}

\section{Lattice correlators and matrix elements}
\label{sec:cal-details}

In this work, we follow the same procedure used to calculate the pion gluon PDF in Sec.~II of Ref.~\cite{Fan:2020cpa,Fan:2021bcr}, following the pseudo-PDF procedure as in Refs.~\cite{Orginos:2017kos,Balitsky:2019krf}.
The gluon operator we used is also the same one as in Eq. 1 in Ref.~\cite{Fan:2021bcr}.
\begin{equation}\label{eq:gluon_operator}
 {\cal O(z)}\equiv\sum_{i\neq z,t}{\cal O}(F^{ti},F^{ti};z)-\sum_{i,j\neq z,t}{\cal O}(F^{ij},F^{ij};z),
\end{equation}
where the operator ${\cal O}(F^{\mu\nu}, F^{\alpha\beta};z) = F^\mu_\nu(z)U(z,0)F^{\alpha}_{\beta}(0)$, $z$ is the Wilson link length.
To extract the ground-state matrix element to construct the reduced pITD defined in Eq.~4, we use a 2-state fit on the two-point correlators and a two-sim fit on the three-point correlators in Eqs.~11 and 12 in Ref.~\cite{ Fan:2021bcr}.

\begin{figure*}[htbp]
\centering
\centering
\includegraphics[width=0.96\textwidth]{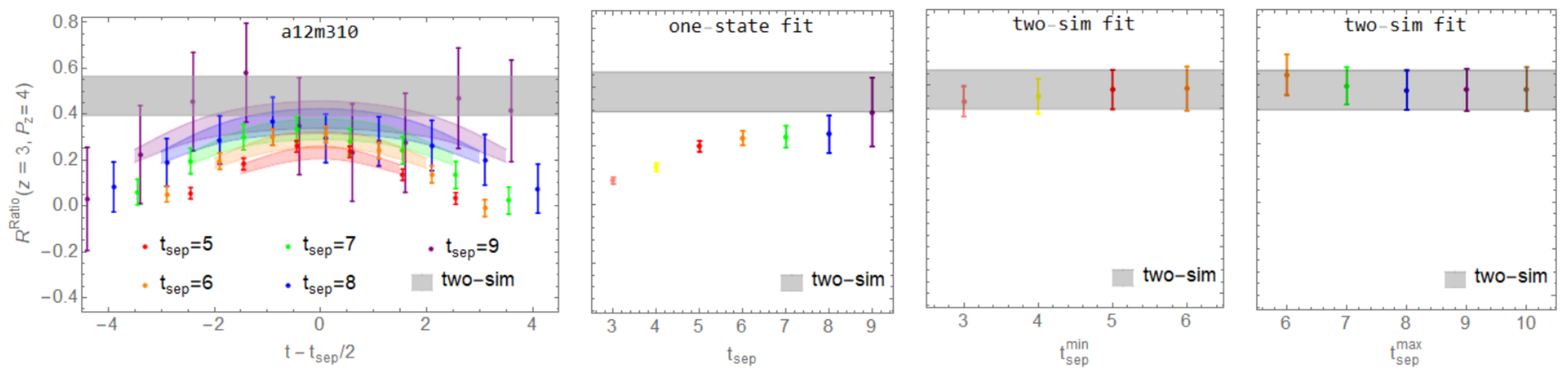}
\includegraphics[width=0.96\textwidth]{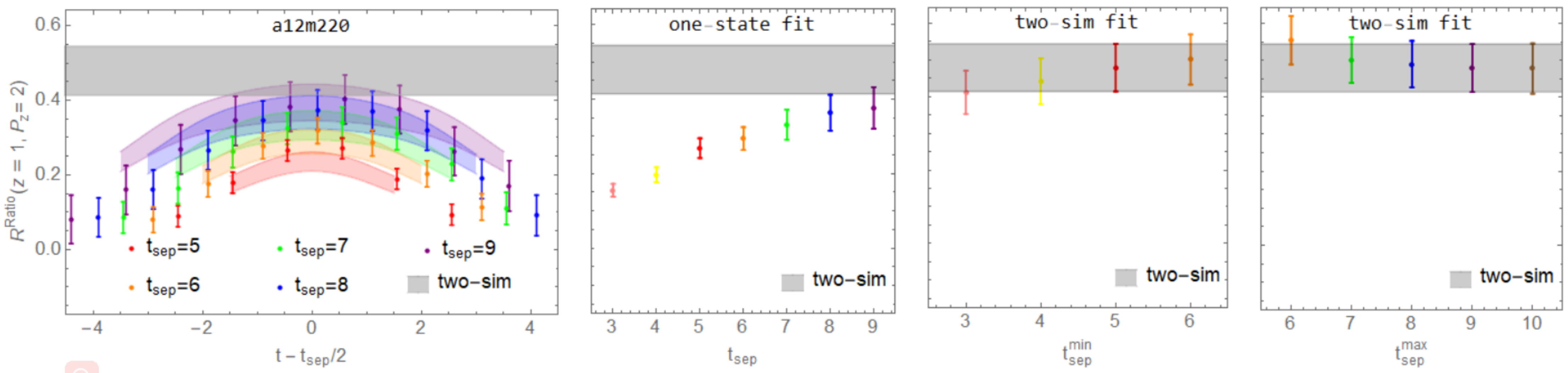}
\caption{Example ratio plots (left), one-state fits (second column) and two-sim fits (last 2 columns) from the preliminary a12m310 light nucleon correlators (upper row) and a12m220 pion correlators~\cite{Fan:2021bcr} (lower row).
The gray band shown on all plots is the extracted ground-state matrix element from the two-sim fit using $t_\text{sep}\in[5,9]$.
From left to right, the columns are:
the ratio of the three-point to two-point correlators with the reconstructed fit bands from the two-sim fit using $t_\text{sep}\in [5,9]$, shown as functions of $t-t_\text{sep}/2$,
the one-state fit results for the three-point correlators at each $t_\text{sep}\in[3,9]$,
the two-sim fit results using $t_\text{sep}\in[t_\text{sep}^\text{min},9]$ as functions of $t_\text{sep}^\text{min}$, and
the two-sim fit results using $t_\text{sep}\in[5,t_\text{sep}^\text{max}]$
as functions of $t_\text{sep}^\text{max}$. }
\label{fig:Ratio-fitcomp}
\end{figure*}

To study the reliability of our fitted matrix-element extraction, we compare to ratios of the three-point to the two-point correlator $R$,
\begin{align}\label{eq:3ptC}
R^{\text{ratio}}_\Phi(z,P_z;t_\text{sep},t)=\frac{C_\Phi^\text{3pt}(z,P_z;t_\text{sep},t)}{C_\Phi^\text{2pt}(P_z;t)}
\end{align}
where the three-point and two-point correlators are defined in Eqs.~11 and 12 in Ref.~\cite{ Fan:2021bcr}.
The left-hand side of Fig.~\ref{fig:Ratio-fitcomp} shows example ratios for the gluon matrix elements from the all ensembles at selected momenta $P_z$ and Wilson-line length $z$.
The ratios increase with increasing source-sink separation $t_\text{sep}$ and the ratios begin to converge at large $t_\text{sep}$, indicating the neglect of excited states becomes less problematic.
The gray bands represent the ground-state matrix elements extracted using the two-sim fit to three-point correlators at five $t_\text{sep}$, where the energies are from the two-state fits of the two-point correlators.
The convergence of the fits that neglect excited states can also be seen in second column of Fig.~\ref{fig:Ratio-fitcomp}, where we compare one-state fits from each source-sink separations:
the one-state fit results increase as $t_\text{sep}$ increases, starting to converge at large $t_\text{sep}$ to the two-sim fit results.
The third and fourth columns of Fig.~\ref{fig:Ratio-fitcomp} show two-sim fits using $t_\text{sep}\in[t_\text{sep}^\text{min},9]$
and
$t_\text{sep}\in[5,t_\text{sep}^\text{max}]$
to study how the two-sim ground-state matrix elements depend on the source-sink separations input into fit.
We observe that the matrix elements are consistent with each other within one standard deviation, showing consistent extraction of the ground-state matrix element, though the statistical errors are larger than those of the one-state fits.
Taking a12m310 ensemble as an example, we observe larger fluctuations in the matrix element extractions when small $t_\text{sep}^\text{min}=3$ and 4, or small $t_\text{sep}^\text{max}=6$ and 7, are used. The ground state matrix element extracted from two-sim fits becomes very stable when $t_\text{sep}^\text{min}>4$ and $t_\text{sep}^\text{max}>7$.

\section{Results and Discussions}
\label{sec:results}
We construct the reduced pITD (RpITD) by taking the double ratio of the pITD as done in the first quark pseudo-PDF calculation~\cite{Orginos:2017kos},
\begin{equation}
\mathscr{M}(\nu,z^2)=
\frac{\mathcal{M}(zP_z,z^2)/\mathcal{M}(0\cdot P_z,0)}{\mathcal{M}(z\cdot 0,z^2)/\mathcal{M}(0\cdot 0,0)}.
\label{eq:RpITD}
\end{equation}

\begin{figure*}[htbp]
\centering
\centering
\includegraphics[width=0.32\textwidth]{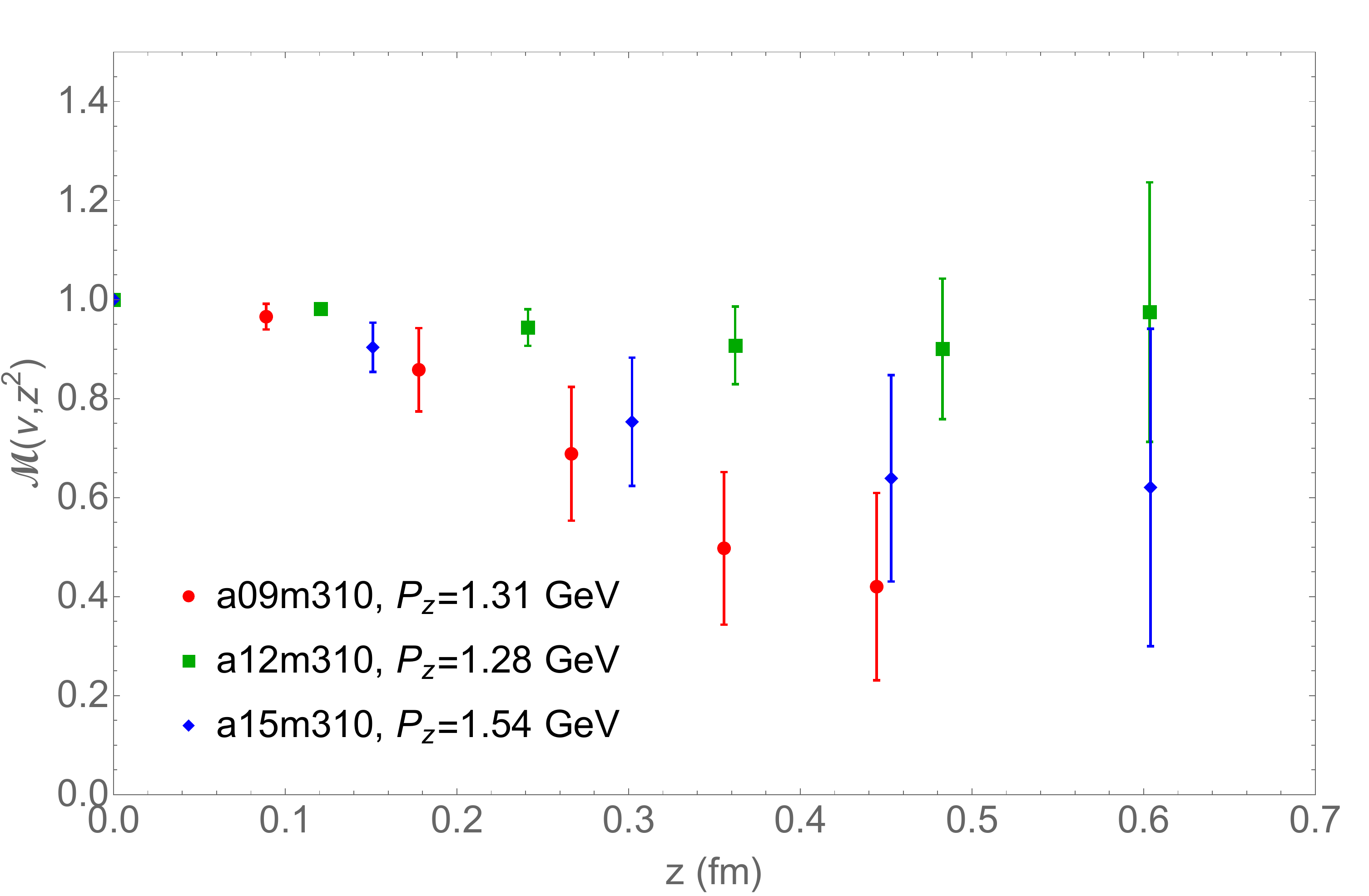}
\includegraphics[width=0.32\textwidth]{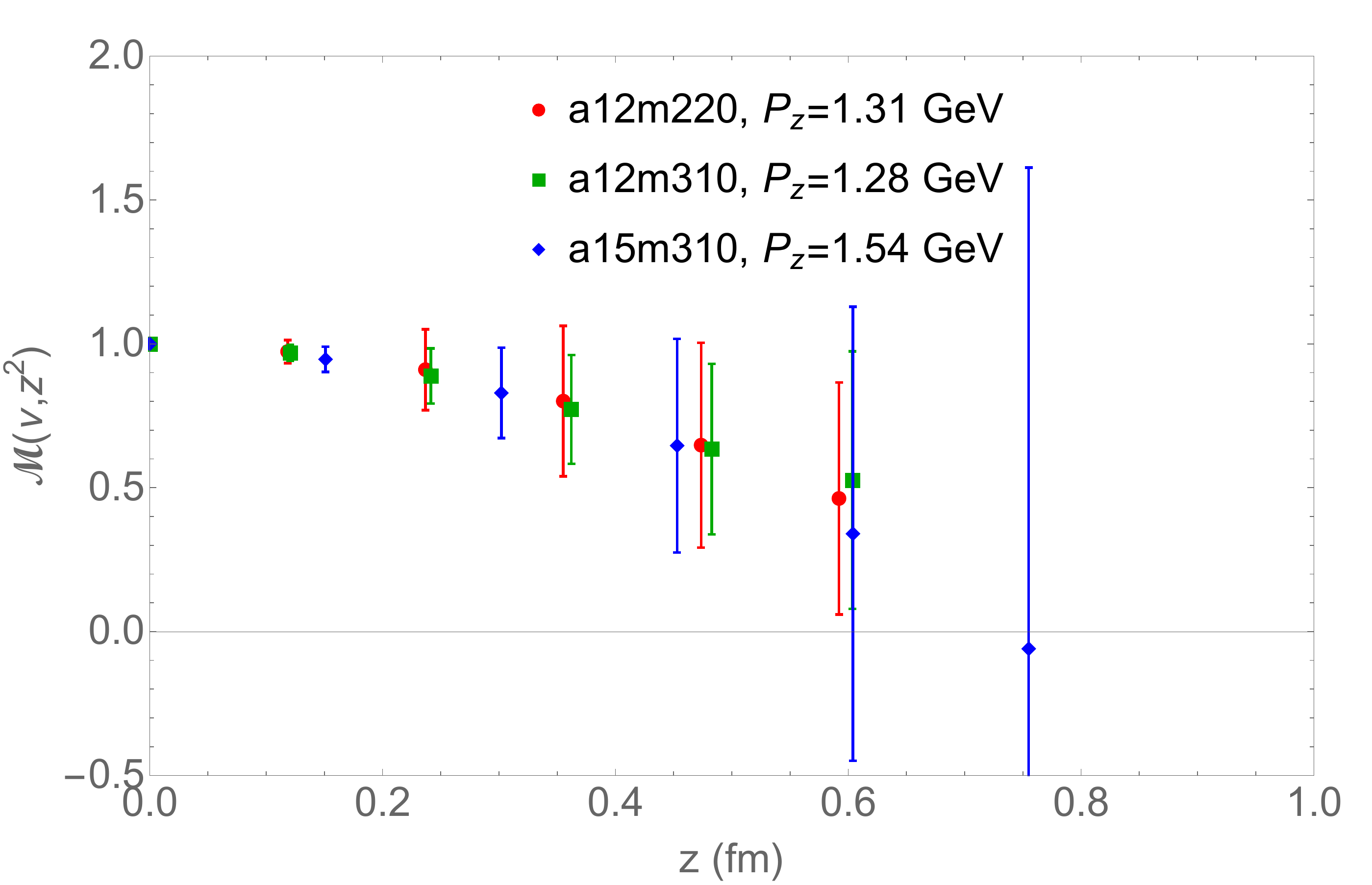}
\includegraphics[width=0.30\textwidth]{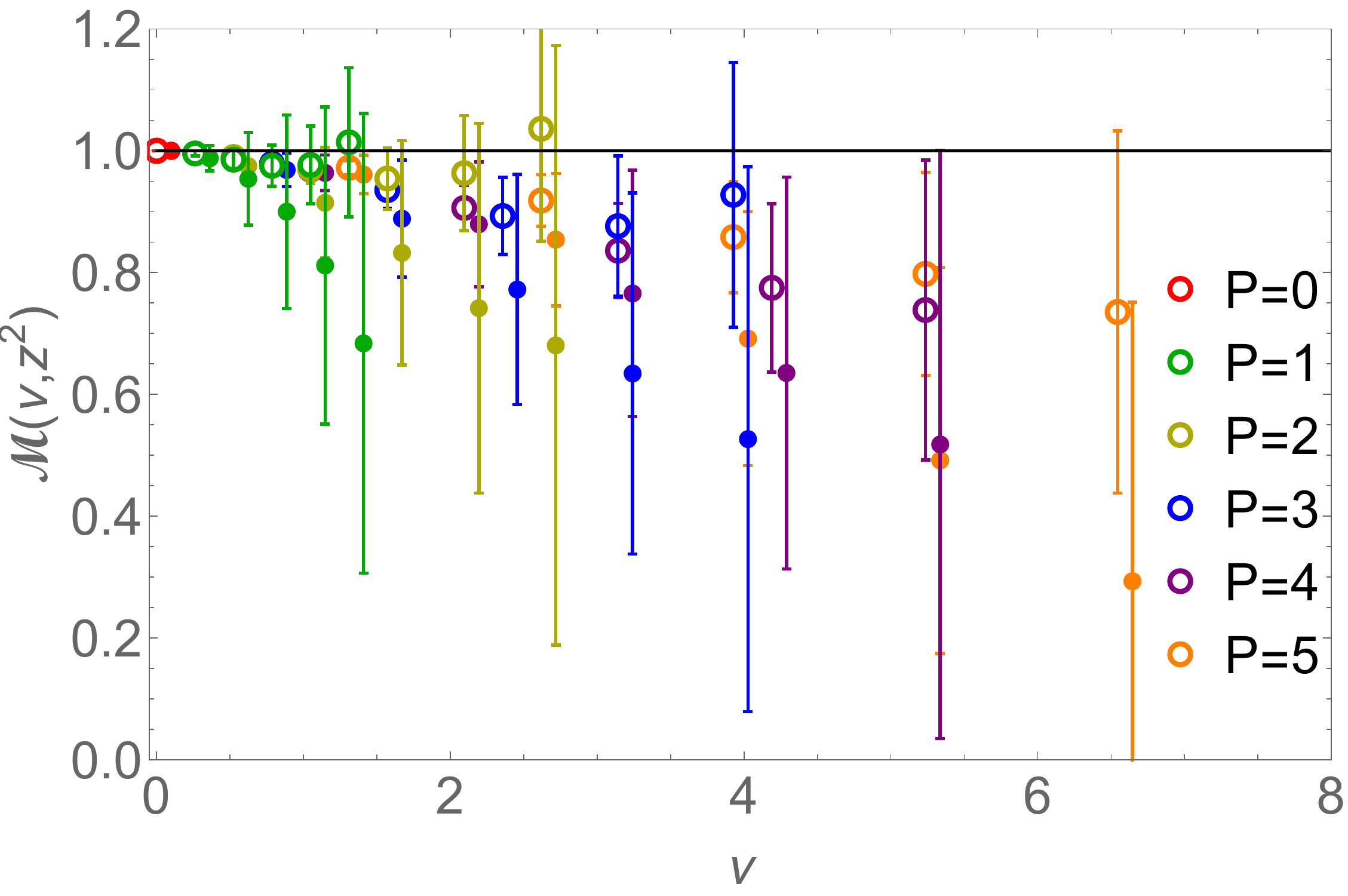}
\caption{Left and middle plots: The RpITDs at boost momentum $P_z \approx 1.3$~GeV for the preliminary nucleon at three different lattice spacings (left), and pion at two different lattice spacings and two different pion masses~\cite{Fan:2021bcr} (middle). They show the lattice-spacing and pion-mass dependence are weak.
Rightmost plot:
The preliminary nucleon and pion RpITDs comparison for a12m310 ensemble. The nucleon RpITD's errors are smaller than pion ones under the same measurements and two-sim fit method.
}
\label{fig:RpITD}
\end{figure*}

The left and middle plots of Fig.~\ref{fig:RpITD} shows the preliminary nucleon and pion RpITDs at boost momentum around 1.3~GeV as functions of the Wilson-line length $z$ for different ensembles. We compare the nucleon RpITDs at three different lattice spacings $a\approx 0.09$, 0.12 and 0.15~fm and the pion RpITDs at two different lattice spacings $a\approx 0.12$ and 0.15~fm and two different pion masses 220 and 310~MeV. We see no noticeable lattice-spacing and pion-mass dependence.

We compare the RpITDs of nucleon and pion for the same ensemble a12m310 with same measurements and two-sim fit method on correlators. The errors of nucleon RpITDs are much smaller than the pion RpITDs and the nucleon RpITDs are larger than the pion ones at large $\nu$, as shown in rightmost plot in Fig.~\ref{fig:RpITD}. The nucleon RpITD values larger than pion ones at larger $\nu$ indicates that the nucleon gluon PDF will approach to 0 quicker than pion gluon PDF as $x\to 1$.

\begin{figure*}[htbp]
\centering
\centering
\includegraphics[width=0.48\textwidth]{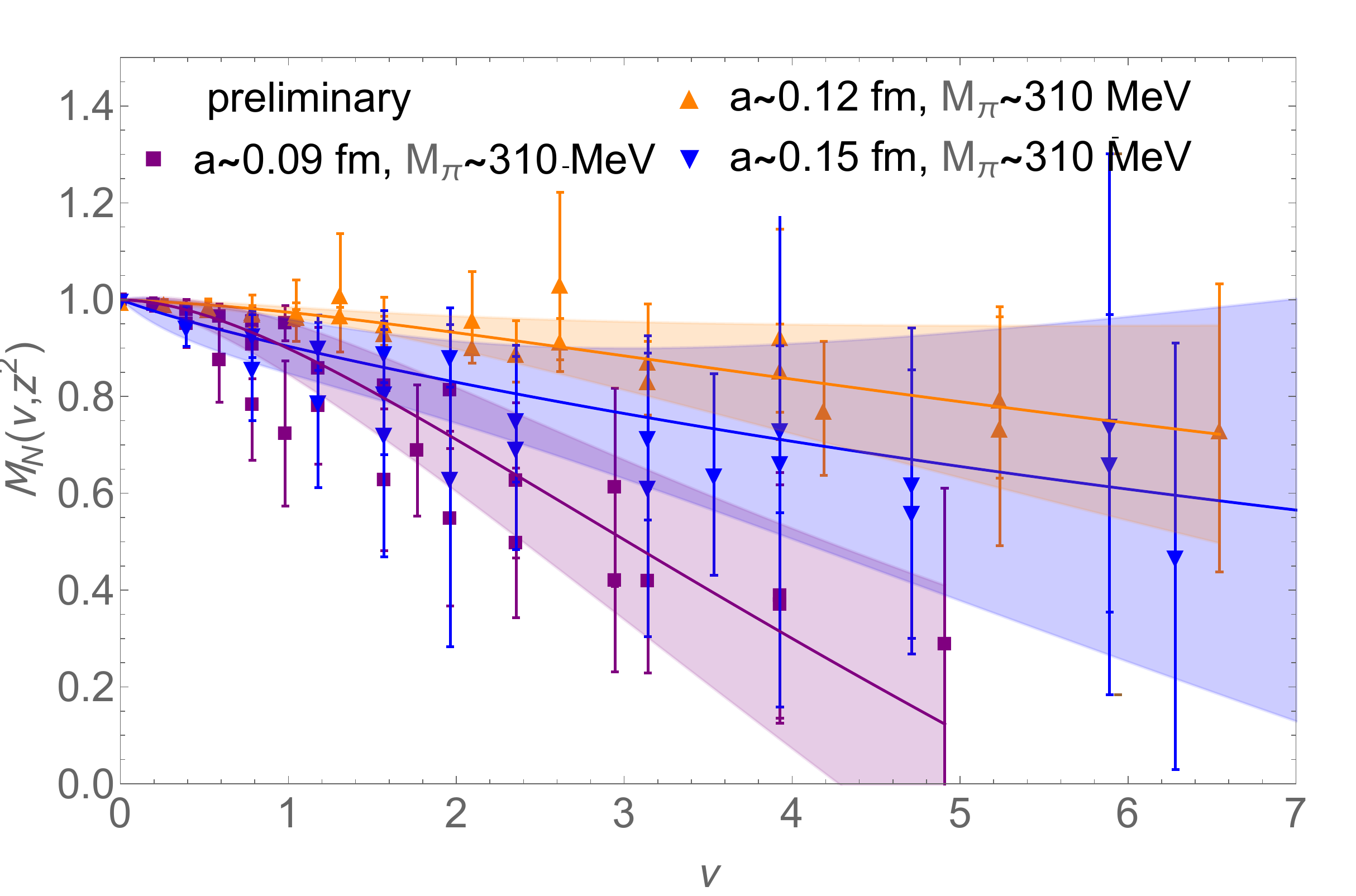}
\includegraphics[width=0.48\textwidth]{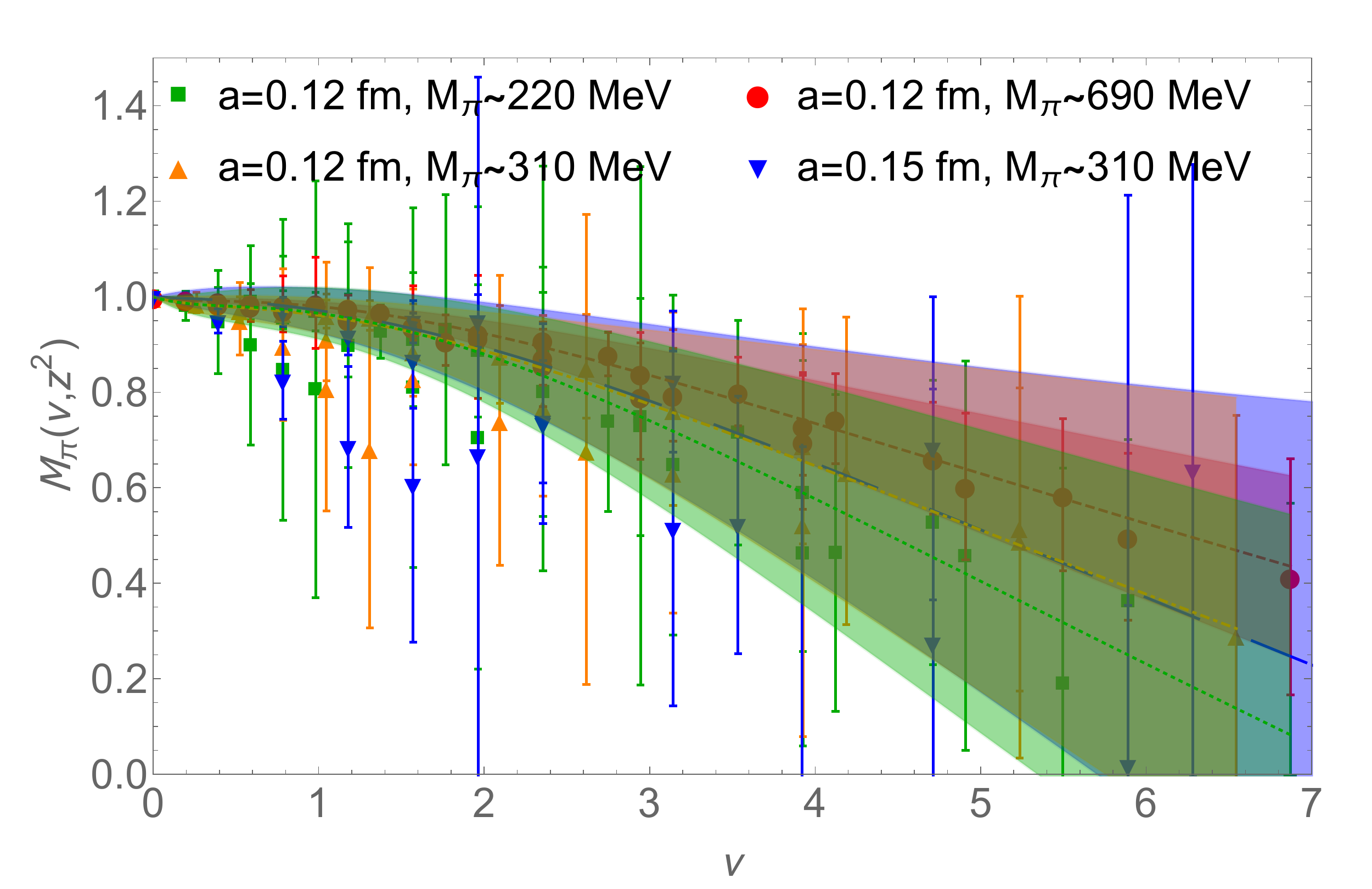}
\caption{
The RpITDs $\mathscr{M}$ with reconstructed bands from ``$z$-expansion'' fits  calculated on different ensembles.
The left plot shows the preliminary nucleon RpITDs and the right plot shows the pion RpITDs.}
\label{fig:RpITD_comp}
\end{figure*}

\begin{figure*}[htbp]
\centering
\centering
\includegraphics[width=0.48\textwidth]{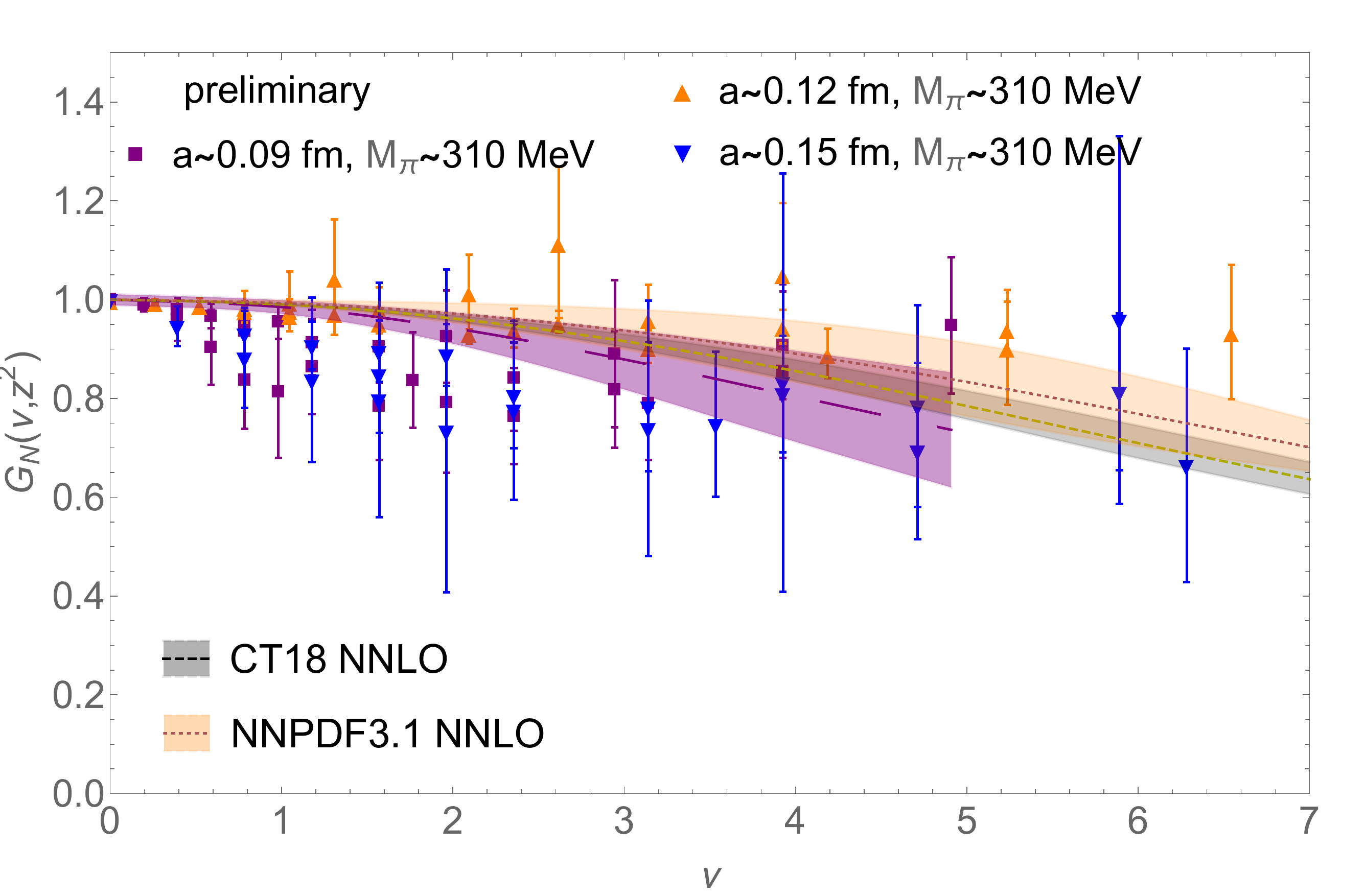}
\includegraphics[width=0.48\textwidth]{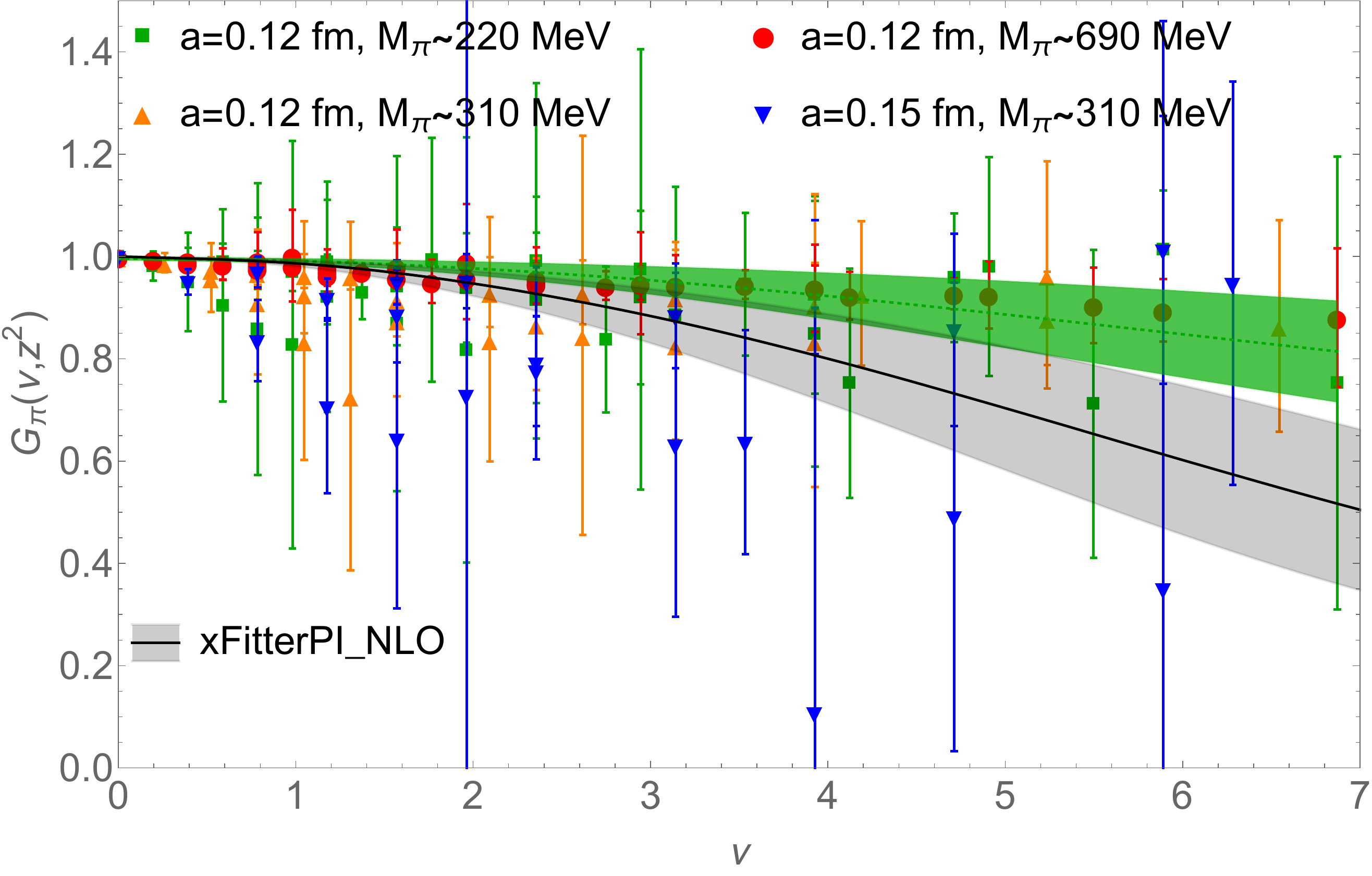}
\caption{The EpITDs $G$ with reconstructed bands from fits on the ensemble with smallest lattice spacing we have for nucleon (left) and the lightest pion mass ensemble for pion (right), comparing with the EpITDs matched from global-fit PDFs. }
\label{fig:EpITD_GF_lat}
\end{figure*}

The evolved pITD (EpITD) $G$ is defined in Eq.~8 in Ref.~\cite{Fan:2021bcr},
\begin{equation}
G(\nu,\mu)=\mathscr{M}(\nu,z^2)
+\int^1_0 dx\, \frac{\alpha_s(\mu)}{2\pi}N_c\ln\left(z^2\mu^2\frac{e^{2\gamma_E+1}}{4}\right)R_B(x)\mathscr{M}(x\nu,z^2).
\label{evolution}
\end{equation}
where $R_B$ is the matching kernel defined in Ref.\cite{Balitsky:2019krf},
$\alpha_s$ is the strong coupling at scale $\mu$,
$N_c=3$ is the number of colors,
and $\gamma_E=0.5772$ is the Euler-Mascheroni constant.
The $z$ dependence of the EpITDs should be compensated by the $\ln{z^2}$ term in the evolution formula.
In principle, the EpITD $G$ is free of $z$ dependence and is connected to the lightcone gluon PDF $g(x,\mu^2)$ through the scheme-conversion
\begin{equation}
G(\nu,\mu)=\int_0^1 dx\frac{xg(x,\mu^2)}{\langle x \rangle_g} (\cos y-\frac{\alpha_s(\mu)}{2\pi}N_c\left( 2R_B(x\nu)+R_L(x\nu)+R_C(x\nu)\right)),
\label{conversion}
\end{equation}
where $R_B, R_L, R_C$ are defined in Ref.\cite{Balitsky:2019krf}.
To obtain EpITDs, we need the RpITD $\mathscr{M}(\nu,z^2)$ to be a continuous function of $\nu$ to evaluate the $x\in[0,1]$ integral in Eq.~\ref{evolution}.
We achieve this by using a ``$z$-expansion'', following the parameters setting and fit strategy that we used in Ref.~\cite{Fan:2021bcr}.
The reconstructed bands from ``$z$-expansion'' on RpITDs of preliminary nucleon and pion are shown in Fig.~\ref{fig:RpITD_comp}.
They describe the RpITD data points well for all ensembles.
EpITDs as functions of $\nu$ on all ensembles for preliminary nucleon and pion are shown in Fig.~\ref{fig:EpITD_GF_lat}.
There are multiple $z$ and $P_z$ combinations for a fixed $\nu$ value at some $\nu$ values.
Therefore, there are points in the same color and symbol overlapping at the same $\nu$ from the same lattice spacing and pion mass.
The EpITDs $G(\nu,\mu)$ should be free of $z^2$ dependence after the evolution.
However, the EpITDs obtained from Eq.~9 in Ref.~\cite{Fan:2021bcr} have $z^2$ dependence from neglecting the gluon-in-quark contribution and higher-order terms in the matching.
The EpITDs also depend on lattice-spacing $a$ and pion-mass $M_\pi$, like RpITDs.
We see that the effects of $a$ and $M_\pi$ dependence on the EpITDs are not large, as shown in the Fig.~\ref{fig:EpITD_GF_lat}.
We also observe a weak dependence on $z^2$ for the RpITDs and EpITDs in Fig.~\ref{fig:EpITD_GF_lat}.
The EpITDs are fitted using the functional form in Eq. 15 in Ref.~\cite{Fan:2021bcr}. The fitted EpITDs are comparing with the EpITDs from the matching of global fit PDFs in Fig.~\ref{fig:EpITD_GF_lat}. They are consistent with each other very well in the small-$\nu$ region and have some deviations at larger $\nu$, because we have fewer points for the EpITDs and they have larger error in the large-$\nu$ region.

\begin{figure*}[htbp]
\centering
\centering
\includegraphics[width=0.46\textwidth]{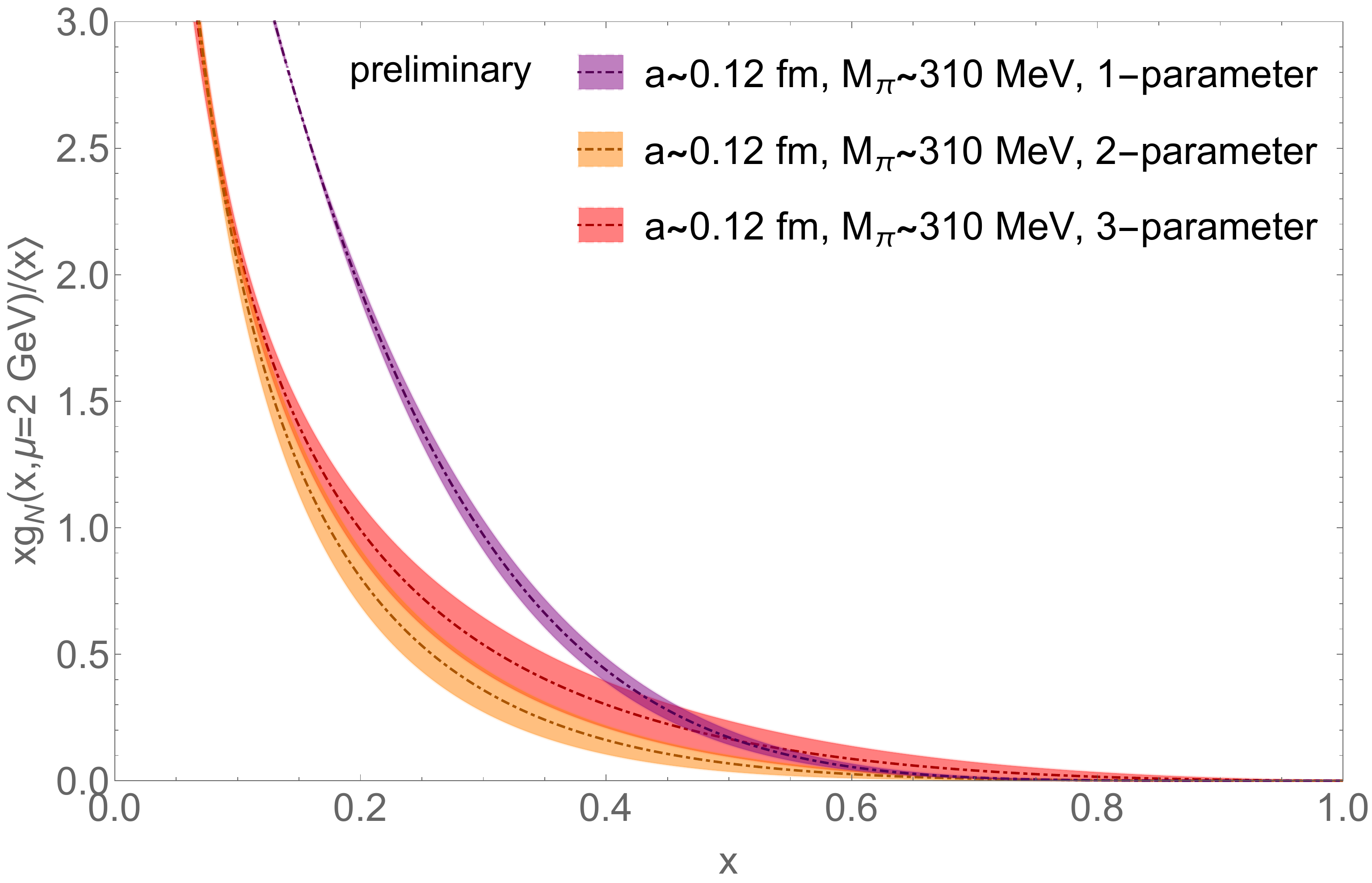}
\includegraphics[width=0.46\textwidth]{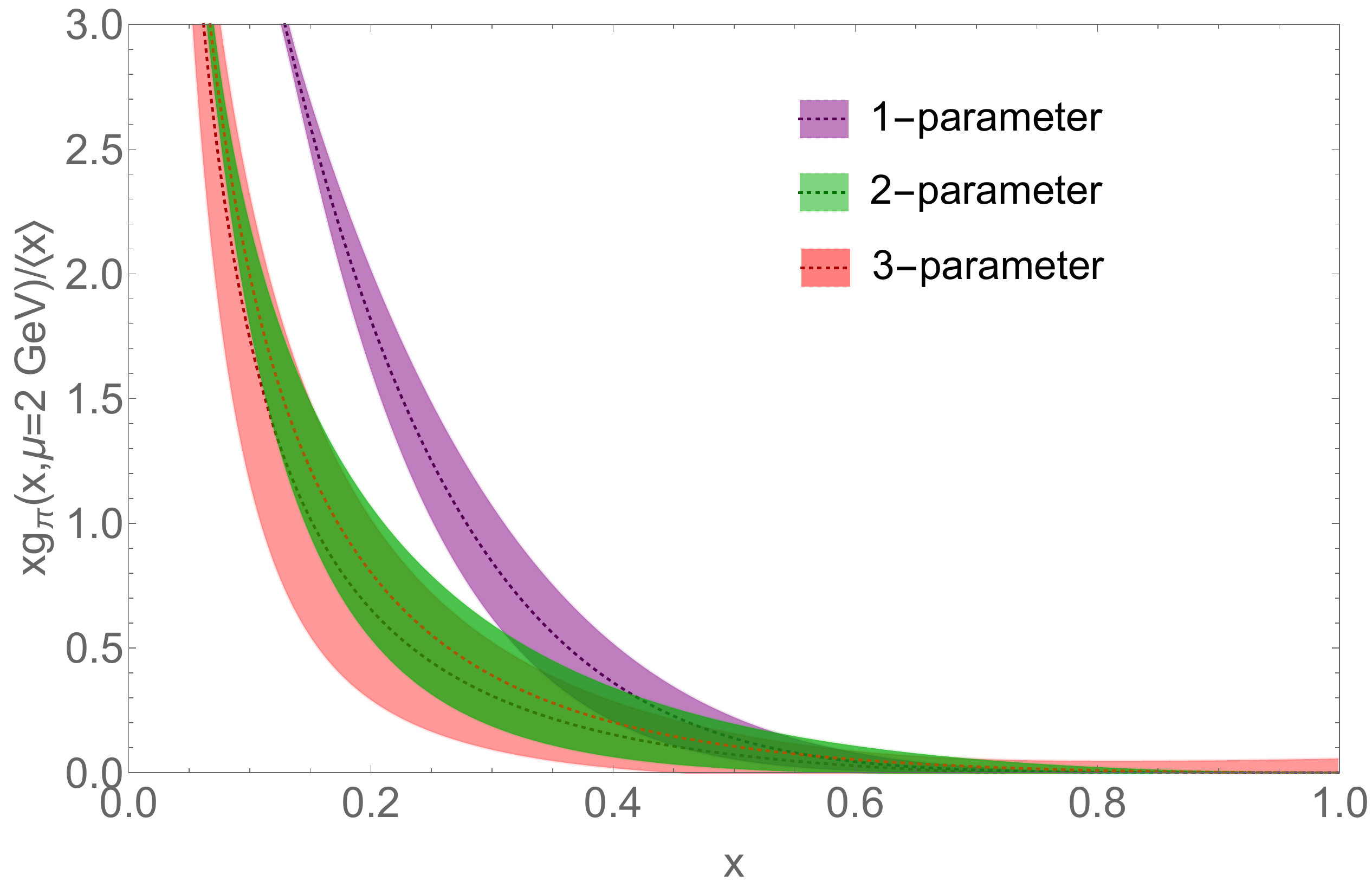}
\caption{The preliminary a12m310 nucleon and a12m220 pion gluon PDF~\cite{Fan:2021bcr} $xg(x, \mu)/\langle x \rangle_g$ at $\mu^2=4\text{ GeV}^2$ as function of $x$ (bottom) calculated with the fitted bands from the 1-, 2- and 3-parameter fits. We conclude that 1-parameter fit on lattice data here is not quite reliable, and the fit results converge for the 2- and 3-parameter fits. }
\label{fig:xgx-para}
\end{figure*}

We investigate the systematic uncertainty introduced by the different parametrization forms which are commonly used for $f_g(x,\mu)$ in PDF global analysis and some lattice calculations.
The first one is the 2-parameter form in $N_0 x^A(1-x)^C$ used in Ref.~\cite{Fan:2021bcr}.
Second, we consider the 1-parameter form $N_1 (1-x)^C$ used in xFitter's analysis~\cite{Novikov:2020snp} (also used in Ref.~\cite{Aurenche:1989sx,Sutton:1991ay}), which is equivalent to the 2-parameter form with $A=0$.
Third, we consider a 3-parameter form $N_3 x^A(1-x)^C(1+D\sqrt{x})$.
We fit the three different forms to the EpITDs of lattice data with $z_\text{max}\approx 0.6$~fm by applying the scheme conversion to the 1-, 2- and 3-parameter PDF forms.
As shown in Fig.~\ref{fig:xgx-para}, there is a big discrepancy between the $f_g(x,\mu)$ fit bands from the 1-parameter fit and the 2-parameter fit in the $x<0.4$ region, but the discrepancy between the 2- and 3-parameter fits is much smaller.
Therefore, we conclude that 1-parameter fit to lattice data here is not quite reliable, and the fit results converge for the 2- and 3-parameter fits.
The same conclusions hold for all other ensembles and pion masses.
Therefore, using the 2-parameter form for our final results is very reasonable.

\begin{figure*}[htbp]
\centering
\centering
\includegraphics[width=0.48\textwidth]{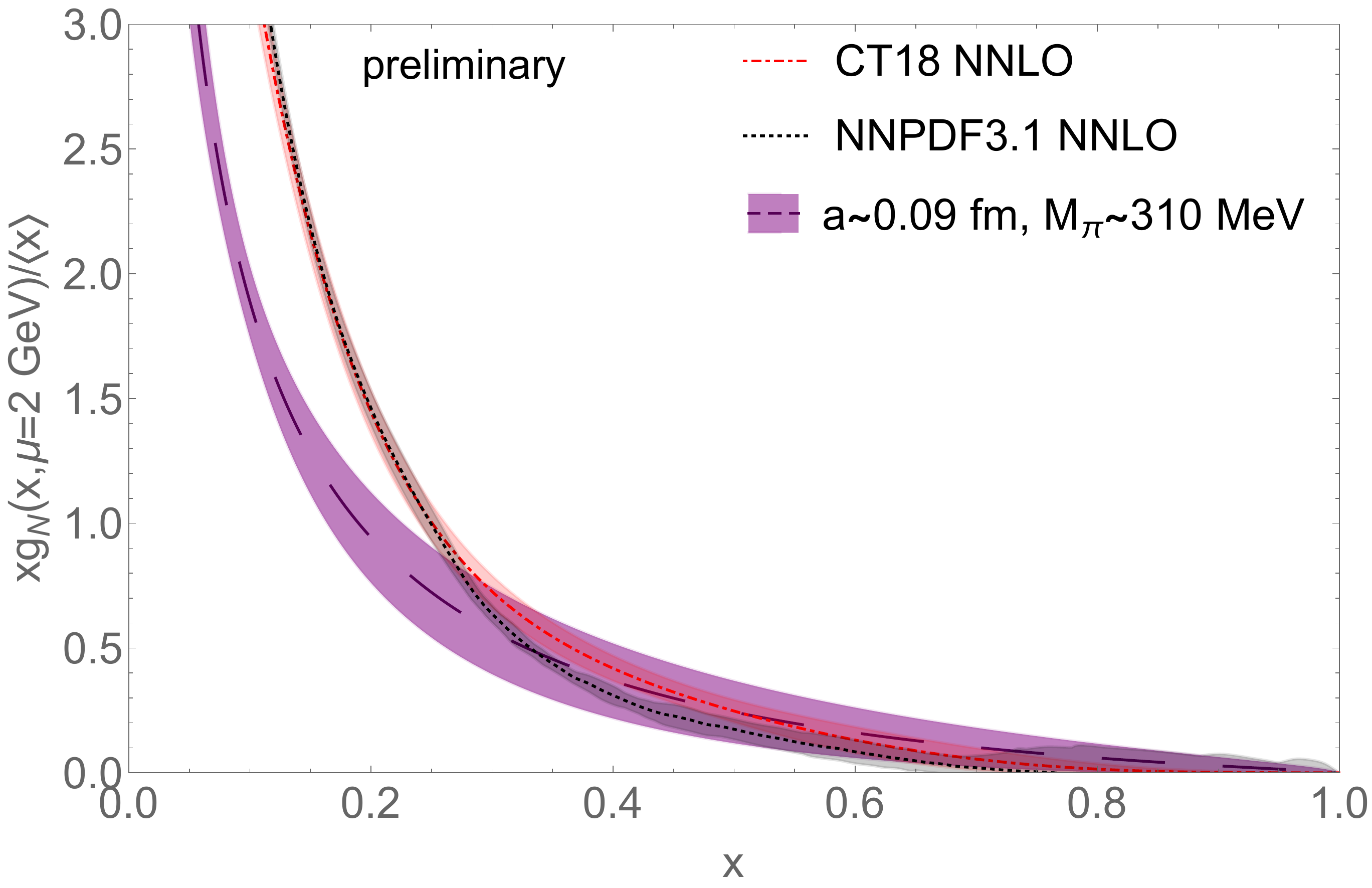}
\includegraphics[width=0.48\textwidth]{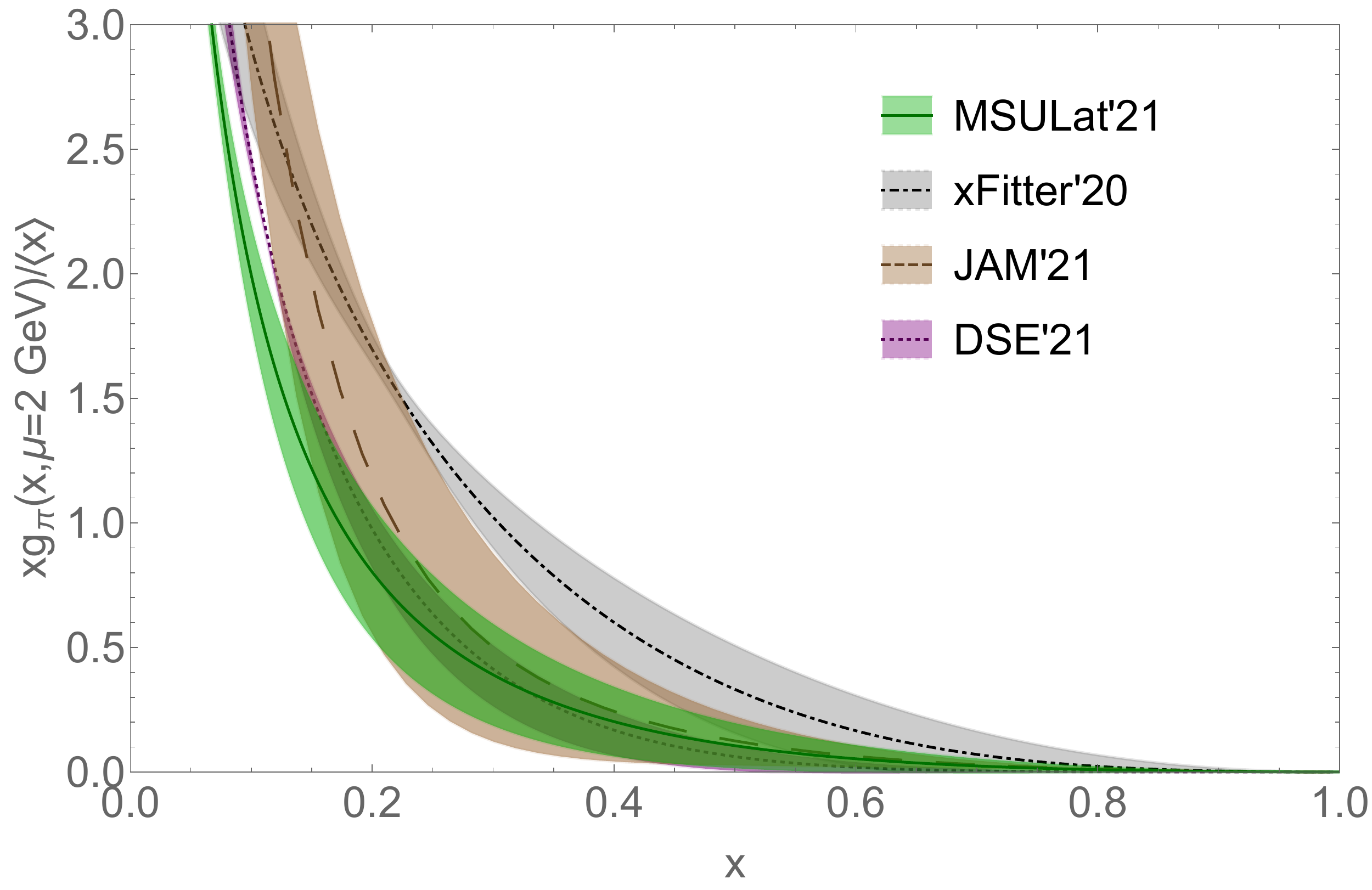}
\includegraphics[width=0.48\textwidth]{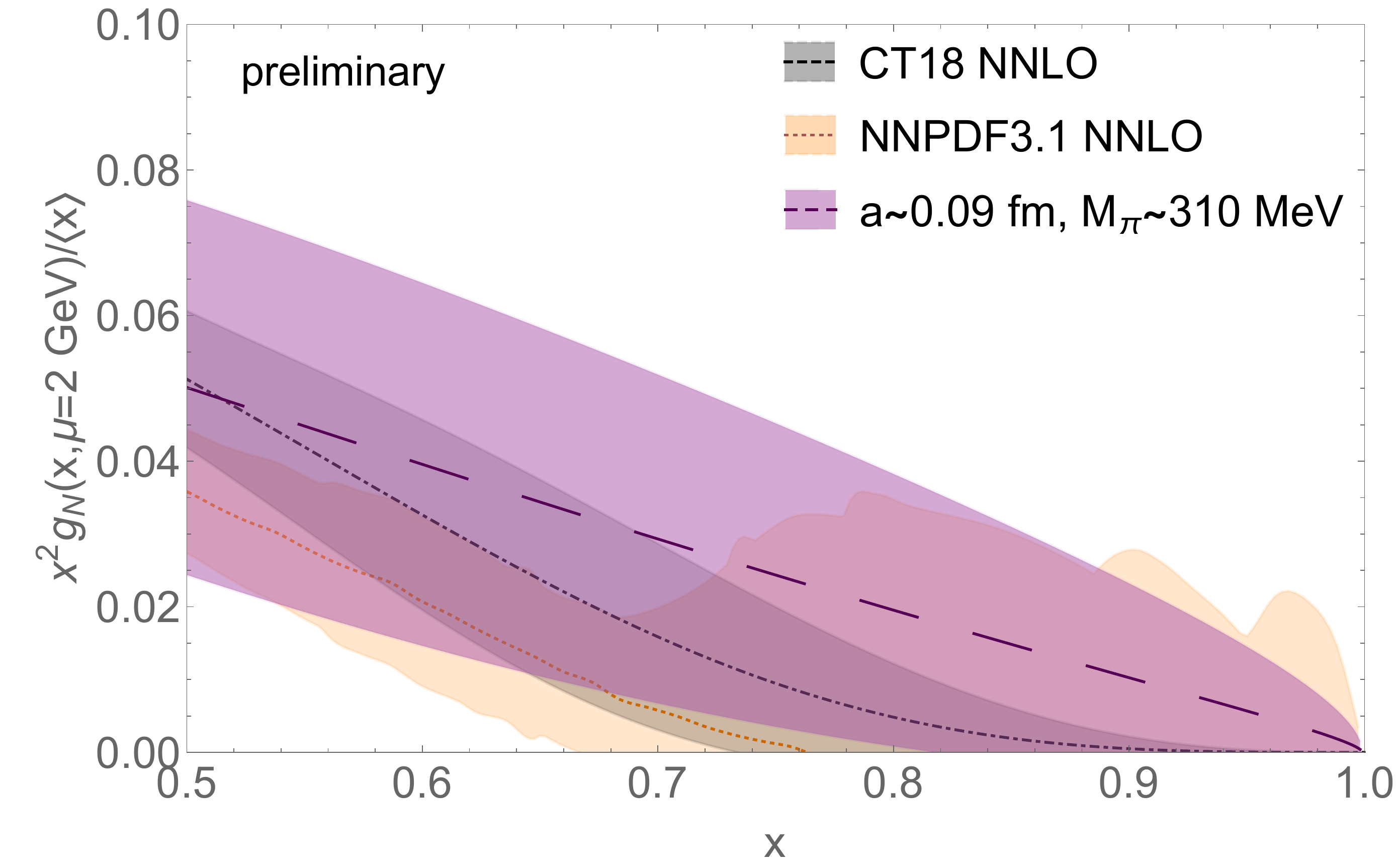}
\includegraphics[width=0.48\textwidth]{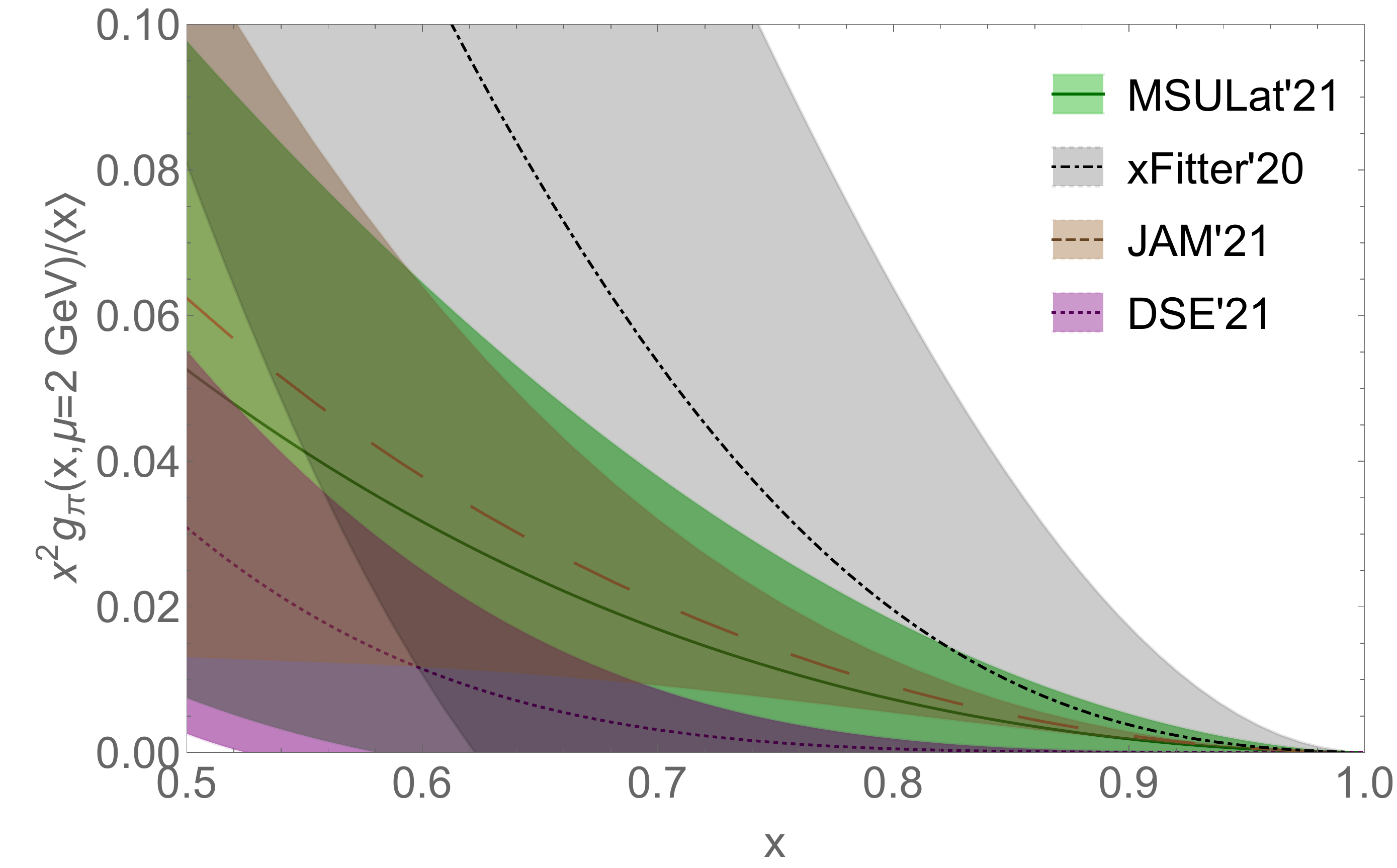}
\caption{The preliminary nucleon (left column) and final pion gluon PDF~\cite{ Fan:2021bcr} (right column) $xg(x, \mu)/\langle x \rangle_g$ (top row) and $x^2g(x, \mu)/\langle x \rangle_g$ (lower row) as a function of $x$ obtained from the functional form fit to the lattice data comparing with the global fit gluon PDFs. Our calculation of gluon PDFs is consistent with global-fit gluon PDFs in the large-$x$ region.
}
\label{fig:xgx-comp}
\end{figure*}

A comparison of our unpolarized preliminary nucleon gluon PDF with CT18 NNLO and NNPDF3.1 NNLO at $\mu=2$~GeV in the $\overline{\text{MS}}$ scheme is shown in left plot in Fig.~\ref{fig:xgx-comp}. We compare our $xg(x,\mu)/\langle x_g \rangle_{\mu^2}$ with the phenomenological curves in the left panel.
We found that our gluon PDF is consistent with the one from CT18 NNLO and NNPDF3.1 NNLO within one sigma in the $x>0.3$ region.
However, in the small-$x$ region ($x< 0.3$), there is a strong deviation between our lattice results and the global fits. This is likely due to the fact that the largest $\nu$ used in this calculation is less than 7, and the errors in large-$\nu$ data increase quickly as $\nu$ increases.
Similarly, the pion gluon PDF comparing with the NLO pion gluon PDFs from xFitter~\cite{Novikov:2020snp} and JAM~\cite{Barry:2018ort,Cao:2021aci} at $\mu^2=4\text{ GeV}^2$ is shown in the right plot in Fig.~\ref{fig:xgx-comp}.
Our pion gluon PDF is consistent with the one from JAM and xFitter NLO PDFs within one sigma in the $x>0.2$ region.
To better see the large-$x$ behavior, we zoom into the large-$x$ region with $x\in[0.5,1]$ and multiply an additional $x$ factor into the fitted $xg(x,\mu)$ for both nucleon and pion gluon PDFs, as shown in the lower row of Fig.~\ref{fig:xgx-comp}. Our large-$x$ results are consistent with global fits over $x \in [0.5,1]$ though with larger errorbars.

\begin{figure*}[htbp]
\centering
\centering
\includegraphics[width=0.48\textwidth]{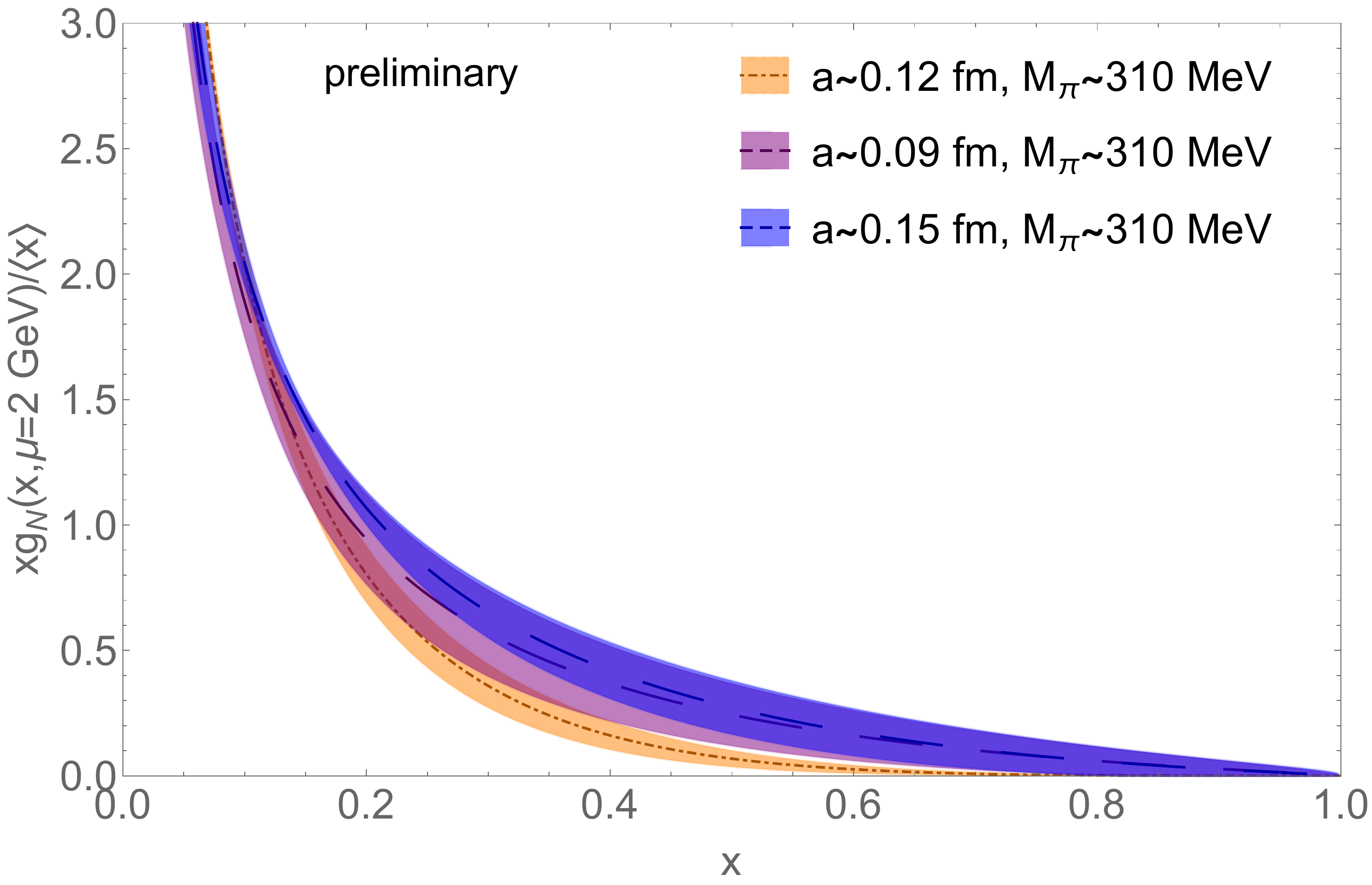}
\includegraphics[width=0.48\textwidth]{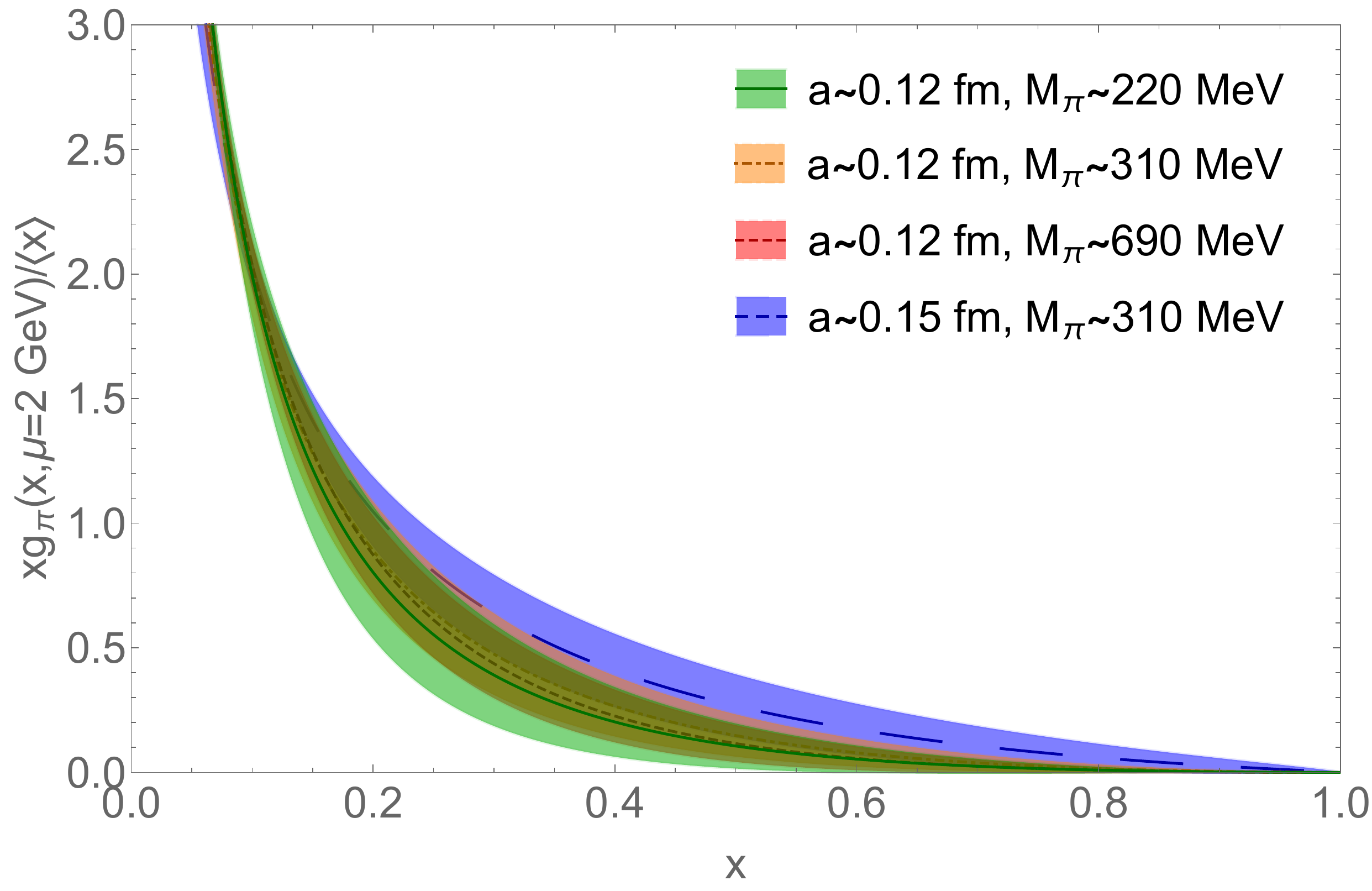}
\caption{The preliminary nucleon and pion gluon PDF~\cite{ Fan:2021bcr} $xg(x, \mu)/\langle x \rangle_g$ as a function of $x$ obtained from the fit to the lattice data from different ensembles with different lattice spacings and pion masses. The fitted gluon PDFs are consistent with each other within one sigma error for all ensembles.}
\label{fig:xgx}
\end{figure*}

\section{Summary}
\label{sec:summary}
We extract the pion and nucleon $x$-dependent gluon PDFs at three pion masses and three lattice spacings and find their dependence to be weak under the current statistics.
We investigated the effects of varying the functional form in the reconstruction fits by using various forms, which are all commonly used or proposed in other PDF works.
We conclude that the 2-parameter fits are sufficient for our current calculation and our final nucleon pion gluon PDF results are presented with the 2-parameter fit results.
There are systematics yet to be studied for the nucleon gluon PDF, such as quark PDF mixing, and the finite $\nu$ extent of the EpITD data.
Thus, in our following work, we will study the the nucleon and pion gluon PDFs with improved statistics and better systematic control.

\section*{Acknowledgments}

We thank MILC Collaboration for sharing the lattices used to perform this study. The LQCD calculations were performed using the Chroma software suite~\cite{Edwards:2004sx}.
This research used resources of the
National Energy Research Scientific Computing Center, a DOE Office of Science User Facility supported by the Office of Science of the U.S. Department of Energy under Contract No. DE-AC02-05CH11231 through ERCAP;
facilities of the USQCD Collaboration, which are funded by the Office of Science of the U.S. Department of Energy,
and supported in part by Michigan State University through computational resources provided by the Institute for Cyber-Enabled Research (iCER).
ZF and HL are partly supported by the US National Science Foundation under grant PHY 1653405 ``CAREER: Constraining Parton Distribution Functions for New-Physics Searches''.

\providecommand{\href}[2]{#2}\begingroup\raggedright\endgroup

\end{document}